\documentclass{article}
\usepackage{graphicx} 
\usepackage{amsmath}
\usepackage{mathtools}
\usepackage{amssymb}
\usepackage{cite}
\usepackage{footnote}
\usepackage[title]{appendix}
\usepackage[scale=0.7]{geometry}
\usepackage{setspace}
\usepackage{hyperref}
\usepackage{float}
\title{Semantic Rate-Distortion Theory with Applications}
\author{Yi-Qun Zhao, Zhi-Ming Ma, Geoffrey Ye Li, Shuai Yuan, Tong Ye, and Chuan Zhou
\footnote{Y. Q. Zhao, Z. M. Ma, S. Yuan, T. Ye, and C. Zhou are with Academy of Mathematics and Systems Science, Chinese Academy of Sciences, Beijing, China (zhaoyiqun@amss.ac.cn, mazm@amt.ac.cn, yuanshuai2020@amss.ac.cn, yetong22@mails.ucas.ac.cn, zhouchuan@amss.ac.cn), and G. Y. Li is with Department of Electrical and Electronic Engineering, Imperial College London, United Kingdom (Geoffrey.Li@imperial.ac.uk). C. Zhou is the corresponding author.}
}
\date{}
\newtheorem{definition}{Definition}
\newtheorem{theorem}{Theorem}

\newtheorem{proposition}{Proposition}

\begin{document}

\maketitle

\begin{abstract}
    Artificial intelligence (AI) is ushering in a new era for communication. As a result, the establishment of a semantic communication framework is putting on the agenda. Based on a realistic semantic communication model, this paper develops a rate-distortion framework for semantic compression. Different from the existing works primarily focusing on decoder-side estimation of intrinsic meaning and ignoring its inherent issues, such as ambiguity and polysemy, we exploit a constraint of conditional semantic probability distortion to effectively capture the essential features of practical semantic exchanges in an AI-assisted communication system. With the help of the methods in rate-distortion-perception theory, we establish a theorem specifying the minimum achievable rate under this semantic constraint and a traditional symbolic constraint and obtain its closed-form limit for a particular semantic scenario. From the experiments in this paper, bounding conditional semantic probability distortion can effectively improve both semantic transmission accuracy and bit-rate efficiency. Our framework bridges information theory and AI, enabling potential applications in bandwidth-efficient semantic-aware networks, enhanced transceiver understanding, and optimized semantic transmission for AI-driven systems.
\end{abstract}

\section{Introduction}

\hspace{1em} The rapid development of modern communication technology has brought the current communication system's symbol transmission rate close to the Shannon limit \cite{01}, while the rise of artificial intelligence has opened up a new path. These phenomena have gradually shifted our focus from the first level of communication, ``How accurately can the symbols of communication be transmitted?" to the second level, ``How precisely do the transmitted symbols convey the desired meaning?" \cite{02}. Therefore, it becomes necessary to establish a semantic framework.


The exploration of semantic communication traces back to the late 1940s and early 1950s. In 1952, a semantic information framework in \cite{03} uses logical probability to measure content significance. Later, in 2011, a model-theoretic framework for semantic communication in \cite{04} extends Shannon’s principles to derive theoretical bounds for lossless semantic compression and reliable transmission under semantic noise. Deep learning enabled semantic communication systems in \cite{28} leverages Transformer architecture to minimize semantic errors and maximize capacity for text transmission, demonstrating remarkable robustness in low-SNR regimes. More recently, a probabilistic model unifying semantic and Shannon frameworks has been developed in \cite{12}, which demonstrates that reliable semantic communication can achieve rates exceeding classical Shannon capacity.

Similar to the traditional communication, semantic compression may be with certain semantic distortion to reduce the required rate, prompting recent investigations into semantic rate-distortion theory. For instance, a framework in \cite{05} jointly encodes semantic information (modeled as latent states) with external observations under dual fidelity metrics, establishing a coding theorem that identifies the minimum achievable rate for a given distortion. A comprehensive analysis of the Gaussian case was subsequently developed in \cite{LSZWH2022}, which was later extended to scenarios involving side information \cite{06} and semantic security constraints \cite{07}. Alternative approaches to semantic compression integrate game-theoretic equilibria with rate-distortion theory \cite{09} or develop rate-distortion frameworks for transmitting learned model distributions \cite{10}.

The rate-distortion-perception (RDP) trade-off, introduced in \cite{13}, provides a comprehensive framework for analyzing data distributions in communication. Initially applied to image restoration, RDP theory demonstrates that controlling statistical divergence improves perceptual authenticity \cite{13}. Subsequent works, such as \cite{14,15,16,17,18,19,20}, formalized this trade-off, deriving information-theoretical limits on coding rates under perceptual constraints. Recent advances include neural compressors approaching theoretical RDP limits \cite{22} and applications to semantic communication. For example, the semantic RDP framework in \cite{11} uses adaptive divergence metrics while an information bottleneck principle in \cite{21} is based on RDP trade-off. More works in this topic can be found in \cite{NBGZWH2025} and the references therein. 

While most of existing works in semantic compression area primarily focus on decoder-side estimation of intrinsic meaning, which often ignore the inherent ambiguity and polysemy issues in semantic interpretation. Probability distributions offer a natural mathematical framework for characterizing such ambiguity and polysemy, as they explicitly model uncertainty. Thus, we argue that greater attention could be directed toward the conditional distribution of semantic information given the observed data rather than pursuing point estimates alone, which is particularly warranted for AI-driven systems. Therefore, the distributions of the transmitted semantic information deserve methodological priority for future intelligent communications.

In this paper, we develop a novel semantic compression framework from the rate-distortion perspective. At its core, 
we introduce a constraint defined through intrinsic semantic probabilities conditioned on extrinsic observations, which both captures practical semantic interactions and addresses the needs of modern AI-driven communications. Using methods from RDP theory, we establish a fundamental coding theorem characterizing the minimum achievable rate for semantic-constrained communication. In particular, the closed-form expression for this fundamental limit is obtained for binary sources.



The rest of this paper is arranged as the following. We will elaborate on our semantic compression framework and raise the semantic rate-distortion trade-off problem in Section \ref{Problem Formulation}. 
Our theoretical contributions will be presented in Section \ref{main results chapter} in the form of main theorems. In Section \ref{binary example}, we will calculate the semantic rate distortion function for a particular semantic scenario and concentrate on the further thinking that this result provokes. Compelling experimental evidence for our semantic compression theory will be demonstrated in Section \ref{experiment}. Finally, we will conclude with a comprehensive synthesis of our findings and their broader influence on the next-generation communication systems in Section \ref{conclusion}.

\section{Problem Formulation}\label{Problem Formulation}

\hspace{1em} Grounded in our theoretical contemplations on semantic information and semantic communication, this section establishes a practical semantic communication model and proposes a semantic distortion measure. 

\subsection{System Model}
\hspace{1em} What is the fundamental difference between a semantic and a traditional communication model? Clearly, unlike the latter that focuses solely on the compression, transmission, and restoration of the original symbols, the semantic communication prioritizes the semantic meaning behind the transmitted symbols. Thus the first crucial question is what exactly is semantics? Or, to put it more bluntly, what is the semantic information 
in a semantic communication system?

Every communication process must correspond to a purpose/goal or multiple ones. For example, in speech transmission, the purpose may be to make the receiver understand the meaning of the speech. If an image is transmitted, the goal could be to identify which kind of animal is in the picture and where the picture was taken. We denote the usage of messages, namely the purpose or goal, as a task T of the communication, which is known to both the sender and the receiver, then the semantics should be the intrinsic information related to task T carried in the symbols of messages.

Yet, any piece of semantic information must be delivered through a symbol string, and any meaningful string must also contain semantics. Therefore, the object we consider in a semantic communication system will shift from individual symbols only to a binary group composed of a piece of semantics and the corresponding symbol string, which we call the pair of intrinsic meaning and extrinsic observation. 

As shown in Figure 1, $s$ is the latent semantic kernel remaining concealed within symbolic representations, while $x$ constitutes the observable medium physically propagating through the communication system. As a result, the semantic coding schemes can only be done on $x$, whereas our attention has changed from the conventional focus on the distortion of symbols to the distortion of semantics hidden in the symbols --- this is the key to semantic communication.

\textbf{Remark:} Why can the semantic coding schemes only be done on $x$, but not on $s$?
\begin{itemize}
    \item In many scenarios, semantic information emerges inherent complexity and ineffability (e.g., molecular structural features of proteins) that defy explicit extraction, necessitating indirect transfer through its extrinsic observation.
    \item According to the data processing inequality, the step of estimating $s$ from $x$ is probably prone to information loss, resulting in semantic distortion.
    \item When the semantics stands as simple and allows reliable estimation, our communication methods are also sufficient for transmitting the extracted semantics (if estimated version of the semantics $\hat{s}=f_1(x)$ and codeword $w_s=f_2(\hat{s})$, then just let semantic encoding function in our framework $f=f_2 \circ f_1$), thereby yielding enhanced generality. 
\end{itemize}

\begin{figure}[t]
\centerline{\includegraphics[width=1\textwidth,trim = 0 0 0 0, clip]{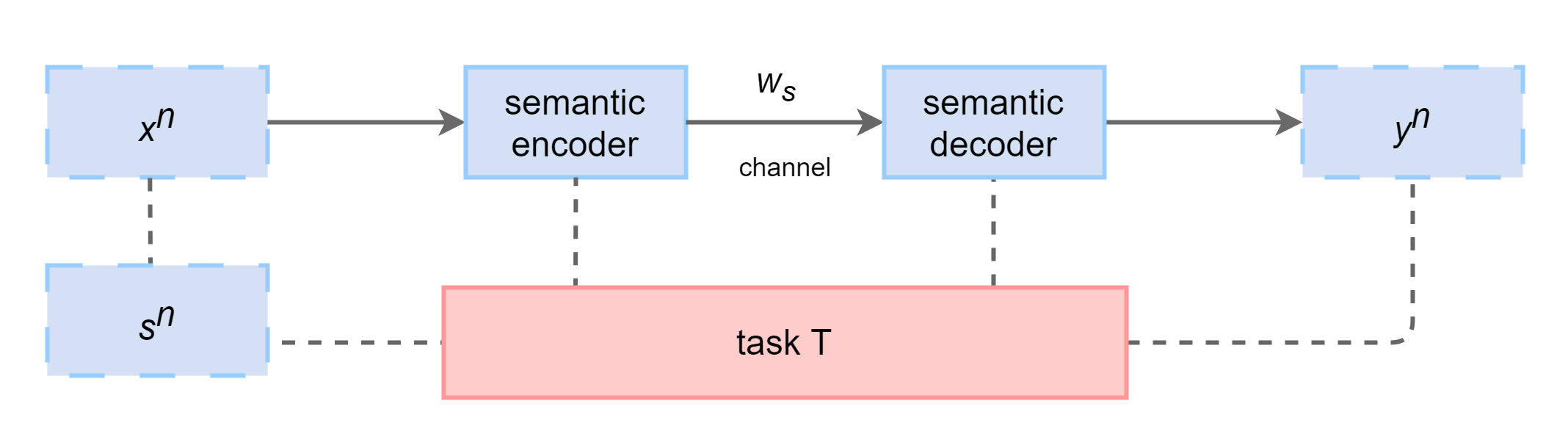}}
\caption{Semantic Communication Model.}
\label{PT}
\end{figure}

Starting from this perspective, the semantic communication procedure can be modeled as follows, as depicted in Figure \ref{PT}:
\begin{itemize}
    \item the $n$-length information pair sent by the semantic source is denoted by $(s^n,x^n)$;
    \item original observation sequence $x^n$ is encoded into a codeword $w_s$ by passing through the semantic encoder;
    \item codeword $w_s$ then enters the channel for transmission;
    \item the semantic decoder decodes the received codeword to obtain reconstructed observation sequence $y^n$;
    \item $y^n$ serves task T.
\end{itemize}

Here are two examples:

\textbf{Example 1:} Task T$_1$ involves transmitting text data and then performing sentiment analysis (also known as opinion mining) on it. The original and reconstructed text data are denoted as $x_1^n$ and $y_1^n$, respectively. Semantic information $s_1^n$ represents the attitudes, emotions, evaluations, or sentiments conveyed by the text data.

\textbf{Example 2:} Task T$_2$ involves transmitting image data and then performing object detection (the identification and localization of specific objects of interest within it) on it. The original and reconstructed image data are denoted as $x_2^n$ and $y_2^n$, respectively. Semantic information $s_2^n$ represents the bounding box coordinates and class labels within the image data.

In this paper, we assume perfect channel and focus on semantic compression.

\textbf{Remark:}
Even if there is no perfect channels in practical engineering, advanced channel coding techniques can achieve near-error-free transmission, thus closely approximating perfect channel conditions. This justifies the rationality of our perfect channel assumption.

\subsection{How to Measure the Distortion of Semantics?}
\hspace{1em} In the semantic communication model in Figure 1, how should we measure the distortion of semantic information? Obviously, the most straightforward way is to compare the (most probable) semantics themselves of both the transmitted symbols and their recovered counterparts. 

However, in practical semantic communication scenarios, a challenge arises when a symbol string manifests referential ambiguity or polysemy --- where a single observation corresponds to multiple plausible semantic interpretations with varying probabilities (e.g.,``orange'' signifies chromatic property or citrus fruit (ambiguity); ``several days'' suggests 3-7 day intervals with equal likelihood (polysemy)). Exclusively transmitting the maximum a posteriori (MAP) semantics risks critical information loss or semantic distortion. At the same time, concurrently in the field of artificial intelligence, modern machine learning systems —-- particularly in classification and object detection tasks —-- map input data to probability-distribution-formatted outputs rather than deterministic predictions.

In consequence, it has become a more reasonable and extensive approach to measure the distortion of semantic information by evaluating the discrepancy between probability distributions of the intrinsic meaning based on the extrinsic observation that is sent and reconstructed.

In Example 1 (2) in Section 2.1, the probability distribution of the intrinsic meaning based on the original and reconstructed extrinsic observation is the probability distribution of various attitudes, emotions, evaluations, or sentiments (bounding box coordinates and class labels) contained in the original and reconstructed text (image) data, respectively.

In certain semantic communication scenarios, especially those with evidentiary demand, such as law enforcement videos transmission for criminal investigation, stringent requirements extend beyond constraining semantic probability distortion to preserve limitations on extrinsic observation fidelity. Hence, our framework incorporates conventional symbol-level distortion as a supplementary criterion.

\subsection{Mathematical Formulation}
\hspace{1em} Let $S$ be a random variable taking values from a semantics set $\mathcal{S}$ with representative element $s$, capturing the intrinsic meaning of a message --- the semantics. For instance, in the practical examples of Sections 2.1 and 2.2, $\mathcal{S}$ in Task T$_1$ is the set of all attitudes, emotions, evaluations, and sentiments and $\mathcal{S}$ in Task T$_2$ is the set of all bounding box coordinates and class labels.

The extrinsic observation of the message at the transmitter and the receiver are then expressed as random variables $X\in\mathcal{X}$ and $Y\in\mathcal{Y}$, respectively, where $\mathcal{X}$ and $\mathcal{Y}$ are two symbols sets. In Example 1, $\mathcal{X}$ is the set of all basic units of original text data, $\mathcal{Y}$ is the set of all basic units of reconstructed text data. In Example 2, $\mathcal{X}$ is the set of all basic units of original image data, $\mathcal{Y}$ is the set of all basic units of reconstructed image data. $x\in\mathcal{X}$ and $y\in\mathcal{Y}$ are specific symbolic realizations, respectively. 

Consider $S$ and $X$ with a known joint probability distribution $p_{S,X}$. $(S_1,X_1)$,\\$\dots$,$(S_n,X_n)$ $i.i.d.$ $\sim  p_{S,X}$, in which $S^n=(S_1,\dots,S_n)$ is an $n$-length intrinsic meaning of messages, and $X^n=(X_1,\dots,X_n)$ models an $n$-length extrinsic observation sequence. 

Conditioned on $x^n$, $S^n$ follows $p_{S^n|x^n}\coloneqq\prod_{i=1}^n p_{S_i|x_i}$, characterizing that the given $n$-length observation, $x^n$, indicates multiple semantic interpretations associated with distinct probability weights in the current context. As outlined in Section 2.2, what we are mainly interested in for semantic communication is messages' precise transmission at the semantic probability level, so incorporating the new constraint --- the probability distribution of the semantics given recovered observation $y^n$ as close as possible to $p_{S^n|x^n}$ --- into the traditional rate-distortion trade-off has become the cornerstone of the semantic rate-distortion problem.

The semantic communication model with dual fidelity metrics in this paper is presented in Figure 2.

\begin{figure}[t]
\centerline{\includegraphics[width=1\textwidth,trim = 0 0 0 0, clip]{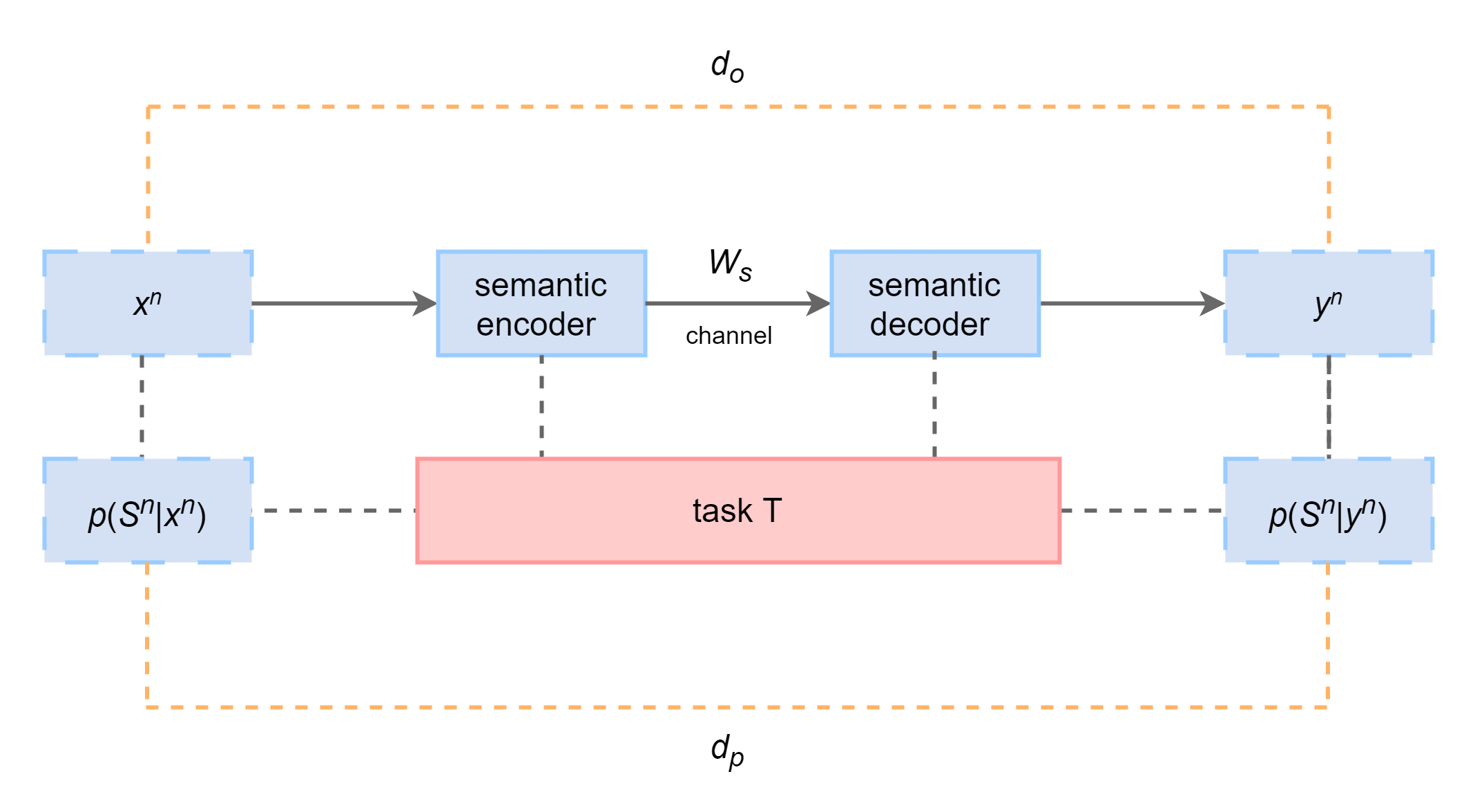}}
\caption{Semantic Communication Model with Dual Fidelity Metrics.}
\end{figure}


The remaining part of this subsection will establish the essential definitions required for the subsequent analysis. Let us start from the definition of the semantic probability distortion for given observation --- the central concern in semantic communication.

\begin{definition} Semantic Probability Distortion Measure $($Based on the Observation$)$: 
A semantic probability distortion measure $($based on the observation$)$ is defined as a function
\begin{equation}
    d_p : \mathcal{P(S)}\times\mathcal{P(S)}\rightarrow\mathbb{R^+},
\end{equation}
where $\mathcal{P(S)}$ denotes the set of all possible probability distributions over semantics set $\mathcal{S}$.
\\Distortion $d_p(p_{S|x}, p_{S|y})$ quantifies the discrepancy between probability distributions of the intrinsic meaning based on extrinsic observations $x$ and $y$. 
\end{definition}

Definition 1 provides a principled measure of semantic distortion between $x$ and $y$. Commonly used metrics in engineering, such as Kullback-Leibler (KL) Divergence and Total Variation (TV) Distance, all satisfy this definition.

Supplementing this, we then introduce the definition of symbolic distortion.

\begin{definition} Observation Distortion Measure: 
An observation distortion measure is defined as a function
\begin{equation}
    d_o : \mathcal{X}\times\mathcal{Y}\rightarrow\mathbb{R^+}.
\end{equation}
Distortion $d_o(x, y)$ quantifies the cost of representing observation $x$ by observation $y$.
\end{definition}

The Observation Distortion Measure defined here subsumes the concept of Shannon's classical distortion measure\cite{01}, covering Hamming distortion, Mean-Squared Error (MSE) distortion, and other practical engineering metrics.

The inter-sequence distortions are rigorously formulated as follows.

\begin{definition} Sequence Distortion Measure: 
The symbolic distortion between observation sequences $x^n$ and $y^n$ is defined as
\begin{equation}
    d_o(x^n,y^n)=\mathop{max}_{i\in\{1,\dots,n\}}\ d_o(x_i,y_i).
\end{equation}
The semantic probability distortion between observation sequences $x^n$ and $y^n$ is defined as
\begin{equation}
    d_p(p_{S^n|x^n},p_{S^n|y^n})=\mathop{max}_{i\in\{1,\dots,n\}}\ d_p(p_{S_i|x_i},p_{S_i|y_i}).
\end{equation}
\end{definition}

Definition 3 specifies that symbolic or semantic distortion between two sequences is measured as the maximum per-component distortion. This approach inherently aligns sequence-level and component-level distortion metrics while providing worst-case performance guarantees.

We now advance to formalize the semantic encoding and decoding process through two mappings.

\begin{definition} Semantic Rate Distortion Code: 
A semantic rate distortion code consists of an $($stochastic$)$ encoding function,
\begin{equation}
    f_n:\mathcal{X}^n\times\mathbb{R}\rightarrow\mathbb{N}^+ ,
\end{equation}
and a $($stochastic$)$ decoding function,
\begin{equation}
    g_n:\mathbb{N}^+\times\mathbb{R}\rightarrow\mathcal{Y}^n.
\end{equation}
\end{definition}

\textbf{Remark:} Both of functions $f_n$ and $g_n$ are endowed with an additional input from $\mathbb{R}$ reflecting the protocol design where the semantic encoder and decoder are permitted to have access to (common, local or hybrid) randomness. 

Then the formal definition of achievability for semantic coding schemes is presented and we leverage this construct to delineate the semantic rate distortion region.

\begin{definition} Achievable Semantic Rate Distortion Triple: A semantic rate distortion triple $(R, D_p,D_o)$ is said to be achievable if there exists a sequence of semantic rate distortion codes $\{f_n, g_n\}$, and a sequence of random variables $\{U_{1,n}\in\mathbb{R},U_{2,n}\in\mathbb{R}\}$ with
\begin{align}
    &\limsup\limits_{n\to+\infty}\frac{H(W_n)}{n} \leq R\label{achievable 1},\\
    &\limsup\limits_{n\to+\infty}E{d}_{p}(p_{S^n|X^n},p_{S^n|Y^n})\leq{D}_{p}\label{achievable 2},\\
    &\limsup\limits_{n\to+\infty}E{d}_{o}(X^n,Y^n)\leq{D}_{o}\label{achievable 3},
\end{align}
where $W_n = f_n(X^n,U_{1,n})$, $Y^n = g_n(W_n,U_{2,n})$, and
\begin{equation}
Ed_p(p_{S^n|X^n},p_{S^n|Y^n}) = E_{(x^n,y^n)\sim p_{X^n,Y^n}}(d_p(p_{S^n|x^n},p_{S^n|y^n}))
\end{equation}
$($or in other words,  
\begin{equation}
Ed_p(p_{S^n|X^n},p_{S^n|Y^n}) = Eh(X^n,Y^n),
\end{equation}
where $h(x^n,y^n)=d_p(p_{S^n|x^n},p_{S^n|y^n}))$.
\end{definition}

\textbf{Remark:} Inequality (\ref{achievable 1}) constrains the asymptotic rate of the semantic rate distortion codes sequence to at most $R$. Simultaneously, inequality (\ref{achievable 2}) and inequality (\ref{achievable 3}) ensure the post-coding semantic and symbolic distortions are asymptotically bounded by $D_p$ and $D_o$, respectively.

\begin{definition} Semantic Rate Distortion Region: 
The semantic rate distortion region for a semantic communication system, which is denoted as $\Omega_s$, is the closure of the set of all achievable semantic rate distortion triples $(R, D_o,D_p)$.
\end{definition}

Building upon the preceding mathematical foundations, we proceed to characterize the rate-distortion trade-off for semantic communication systems through operational and informational perspectives and prove their equivalence in the following section.

\begin{definition} Operation Semantic Rate Distortion Function: 
The operation semantic rate distortion function, $R^O({D}_{p},{{D}_{o}})$, for a semantic communication system is defined as
\begin{align}
    R^O({D}_{p},{{D}_{o}})
    =\mathop{inf}_{R}\{R:(R,{D}_{p},{{D}_{o}})\in\Omega_s\}.
\end{align}
\end{definition}

The operation definition is intuitive: it characterizes minimum achievable rate $R$ for semantic rate distortion codes sequences adhering to asymptotic distortion bounds $D_p$ (semantic) and $D_o$ (symbolic). However, deriving this minimum rate under said definition proves computationally intractable. We therefore pursue an alternative approach, which is consistent with Shannon's conventional way.

\begin{definition} Information Semantic Rate Distortion Function: 
The information semantic rate distortion function, $R^I({D}_{p},{{D}_{o}})$, for a semantic communication system with distortion measures $d_p(\cdot, \cdot)$ and $d_o(\cdot, \cdot)$ is defined as
\begin{equation}\label{definition RI}
    R^I({D}_{p},{D}_{o})=\mathop{min}_{ \substack{p_{Y|X}:E{d}_{p}({p}_{S|X},{p}_{S|Y})\leq{D}_{p}\\E{d}_{o}(X,Y)\leq{D}_{o}}}  {I(X;Y)},
\end{equation}
where $p_{S,X,Y}=p_{Y|X}p_{S,X}$ $($that is to say, $Y$ and $S$ are independent given $X$$)$.
\end{definition}

The information semantic rate distortion function is defined via a tractable optimization problem. By establishing its equivalence to the operation definition, we can determine the fundamental limit of the rate-distortion theory for semantic communication: the minimum achievable rate of semantic rate distortion codes sequences meeting prescribed semantic ($D_p$) and symbolic ($D_o$) distortion constraints is obtained.

\section{Semantic Rate-Distortion Theory}\label{main results chapter}

\hspace{1em} We commence our analysis by investigating the basic property of the information semantic rate distortion function, $R^I({D}_{p},{{D}_{o}})$, for succeeding equivalence proof.

As $D_p $ (or $D_o$) increases, the feasible region of \eqref{definition RI} also becomes larger, which implies that $R^I(D_p,D_o)$ is decreasing. Therefore, we have Proposition 1.

\begin{proposition} Separate Monotonic Decreasing Property\label{SMDP}: 
The information semantic rate distortion function, $R^I(D_p,D_o)$, is monotonically decreasing in each variable, which refers to the fact that $R^I(D_p,D_o)$ is a decreasing function of $D_o$ for each fixed $D_p$ and $R^I(D_p,D_o)$ is decreasing with respect to $D_p$ for each fixed $D_o$.
\end{proposition}

We first show the information semantic rate distortion function is not less than the operation one.

\begin{theorem}\label{THA}
\begin{equation}
    R^I({D}_{p},{{D}_{o}})\geq R^O({D}_{p},{{D}_{o}}).
\end{equation}
\end{theorem}

Following an approach similar to \cite[Theorem 1]{15}, we prove Theorem 1 by explicitly constructing a sequence of length-$n$ semantic codes satisfying that 1) they achieve distortions $(D_o,D_p)$, and 2) their rates converge to $R^I({D}_{p},{{D}_{o}})$ as $n\rightarrow\infty$. The codes construction relies crucially on the Poisson representation lemma \cite{proof}, which provides a framework for designing codes with the required properties. The details on the proof of Theorem 1 can be found in Appendix A.

We now show that the converse maintains correct under a relatively non-restrictive condition.

\begin{theorem}\label{converse part}
If the information semantic rate distortion function, $R^I({D}_{p},{{D}_{o}})$, is lower semicontinuous, i.e.,
\begin{equation}
\liminf\limits_{(D_p,D_o)\rightarrow(P,D)}R^I({D}_{p},{{D}_{o}}) \geq R^I(P,D),\ \  \forall P\geq 0 ,  D\geq 0,
\end{equation}
then $R^O(D_p,D_o) \geq  R^I(D_p,D_o)$, and further we have $R^O(D_p,D_o)=R^I(D_p,D_o)$ in conjunction with Theorem 1.
\end{theorem}

The proof of Theorem 2 adopts the standard converse arguments developed for the rate distortion function \cite[p.317]{Cover2006} and subsequently applied to the rate distortion perception function \cite{14}. The details on the proof of Theorem 2 can be found in Appendix B.

\textbf{Remark:} It can be demonstrated that the information semantic rate distortion function, $R^I(D_p,D_o)$, is lower semicontinuous for most circumstances. For instance, we are going to show later in Proposition 2 that when sets $\mathcal{S},\mathcal{X}$ and $\mathcal{Y}$ are finite, then $R^I(D_p,D_o)$ is lower semicontinuous for most widely used $d_p(\cdot,\cdot)$, including Total Variation (TV) Distance, Weierstrass Distance and all $f$-divergence. As long as $R^I(D_p,D_o)$ is lower semicontinuous, we must have $R^O(D_p,D_o)=R^I(D_p,D_o)$, which 
points out that the operational and informational minimum rates are consistent, indicating the equivalence of the two definitions of the semantic rate distortion function. We will omit its superscript $O$ or $I$ when $R^O(D_p,D_o)=R^I(D_p,D_o)$ for convenience.

The following proposition, proved in Appendix C, provides a sufficient condition for the lower semicontinuity of $R^I(D_p,D_o)$. 

\begin{proposition} Lower Semicontinuity\label{Lower semicontinuity}: 
    Suppose sets $\mathcal{S}, \mathcal{X}$ and $\mathcal{Y}$ are finite, distortion measure $d_p(\cdot,\cdot)$ is continuous with respect to its second argument, and for any $y\in\mathcal{Y}$ and any $q_{Y|X}\in S_y$ with $S_y=\{p_{Y|X} : p_Y(y) = 0\}$, it holds that
\begin{equation}\label{regular condition}
\lim\limits_{\substack{p_{Y|X}\rightarrow q_{Y|X}\\p_{Y|X}\notin S_y}} \sum\limits_{x\in\mathcal{X}}p_{Y|X}(y|x)p_X(x)d_p(p_{S|x},p_{S|y}) = 0,
\end{equation}
then the information semantic rate distortion function, $R^I(D_p,D_o)$, is lower semicontinuous over $\mathbb{R^+}\times\mathbb{R^+}$.
\end{proposition}

\textbf{Remark:} Identity \eqref{regular condition} is a regular condition that guarantees the continuity of function $E{d_p(p_{S|X},p_{S|Y})}$. 
Note that if $d_p(\cdot,\cdot)$ is bounded, then identity \eqref{regular condition} holds, implying TV Distance and all Weierstrass Distance satisfy \eqref{regular condition}.
Furthermore, it can be easily verified that all $f$-divergence also meets \eqref{regular condition}. Consequently, the applicability of Theorem 2 extends to most practical settings.

\section{Semantic Rate Distortion Function for Binary Case}\label{binary example}

\hspace{1em} Consider a simple yet realistic semantic scenario and derive an explicit expression of the semantic rate distortion function to get insight. Suppose a factory device has two states: operational and faulty. This device transmits binary signals (0 or 1) to indicate its state at any given time. Due to the imperfect reliability of its transmission mechanism, there is a certain probability of it sending either 0 or 1 in both states. We model the device’s state at a given moment using a semantic random variable $S$, whose value set $\mathcal{S}=\{0, 1\}$ represents operational and faulty states, respectively. The transmitted symbol is modeled by a symbolic random variable $X$, which takes values from $\mathcal{X}=\{0, 1\}$. Random variable $Y$ represents the symbol received by the console, and $Y \in \mathcal{Y} = \{0,1\}$. According to statistical phenomenon, the joint distribution of $S$ and $X$ satisfies 
\begin{align}
&p_S(0)  = 1 - p_S(1) = \rho,  \label{p_S}\\
&p_{X|S} =  \begin{bmatrix}q_1& 1-q_1\\1-q_2 & q_2 \end{bmatrix}, \label{p_X_given_S}
\end{align}
where $\rho,q_1,q_2\in[0,1]$.
The closed-form expression of the semantic rate distortion function for this binary case is supplied in this section. Although trivialized for real-world deployment, it offers an enlightening example that facilitates foundational insights into the dynamics of the semantic rate distortion function.

We adopt TV Distance $d_{\text{TV}}$ as semantic probability distortion measure $d_p(\cdot,\cdot)$ and Hamming Distance $d_H$ as observation distortion measure $d_o(\cdot,\cdot)$. The truth derived directly from Proposition \ref{Lower semicontinuity} and Theorem \ref{converse part} is that the information semantic rate distortion function, $R^I(D_p,D_o)$, is lower semicontinuous under this setting and $R^I(D_p,D_o) = R^O(D_p,D_o)$, confirming that $R^I(D_p,D_o)$ is the minimum achievable rate for the binary case (superscript $I$ will be then omitted for convenience).

Theorem \ref{binary-case-closed-form}, proved in Appendix D, exhibits the closed-form expression of $R(D_p,D_o)$ when $(S,X)$ follows a doubly symmetric binary distribution with $\rho = 0.5$ and $q_1 = q_2 = q$.
\begin{theorem}\label{binary-case-closed-form}
  Let $\mathcal{S} = \mathcal{X} = \mathcal{Y} = \{0,1\}$, and $(S,X)$ follows joint distribution \eqref{p_S} and \eqref{p_X_given_S} with $\rho = 0.5$ and $q_1 = q_2 = q$. The solution to the optimization problem in \eqref{definition RI} with $d_p = d_{\text{TV}}$ and $d_o = d_H$ for $D_p\in[0,1]$ and $D_o\in[0,1]$ is
\begin{equation}\label{closed form binary case}
R(D_p,D_o) = \left\{\begin{aligned} & 1 - h_2\left(\frac{1 - \sqrt{1 - \frac{2D_p}{C}}}{2}\right), & D_p\in[0,a(D_o)],\\
& 1 - h_2\left(\min\left\{D_o,\frac{1}{2}\right\}\right), & D_p\in(a(D_o),1],\end{aligned}\right.
\end{equation}
where $h_2(x) = -x\log x - (1-x)\log (1-x)$ is the binary entropy function, $C = |1-2q|$, and
\begin{equation}\label{aD_o}
a(D_o) = \left\{\begin{aligned} & 2CD_o(1-D_o),&0\leq D_o \leq \frac{1}{2},\\& \frac{C}{2},&\frac{1}{2} < D_o \leq 1.\end{aligned}\right.
\end{equation}
\end{theorem}

\begin{figure}[htbp]
\centerline{\includegraphics[width=0.5\textwidth,trim = 20 13 16 12, clip]{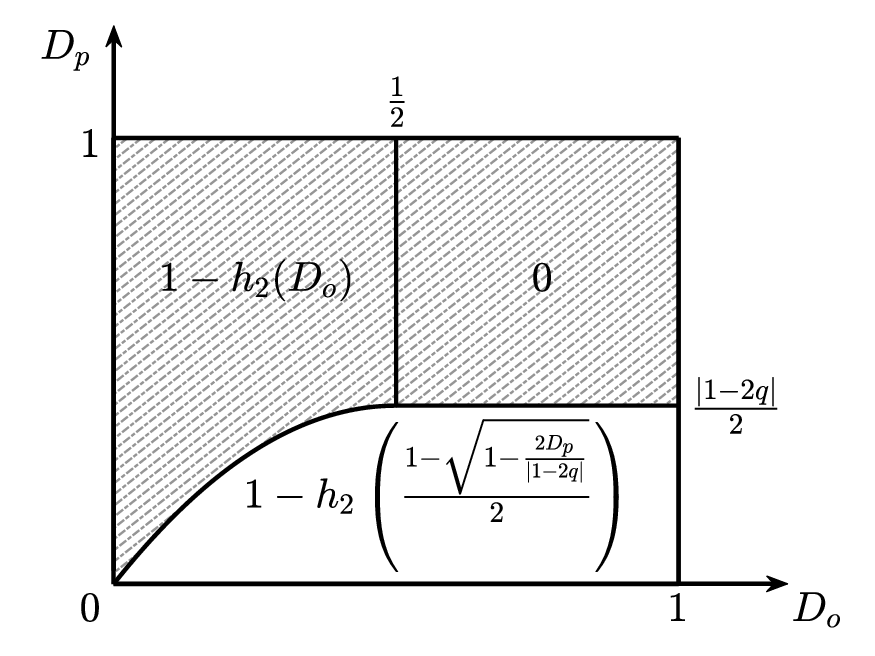}}
\caption{The Closed-Form Expression of $R(D_p,D_o)$ for the Binary Case.}
\label{binary-case}
\end{figure}

Figure \ref{binary-case} provides the visualization of the closed-form expression in \eqref{closed form binary case}. Notably, for small $D_p$, $R(D_p,D_o)$ is governed by the semantic probability distribution constraint based on the observation, while for larger values of $D_p$, the function depends only on $D_o$ and degenerates to the traditional rate distortion function for symmetric binary sources, as in the shaded area in Figure 2.

\begin{figure}[H]
\centerline{\includegraphics[width=1.0\textwidth,trim = 103 16 120 21, clip]{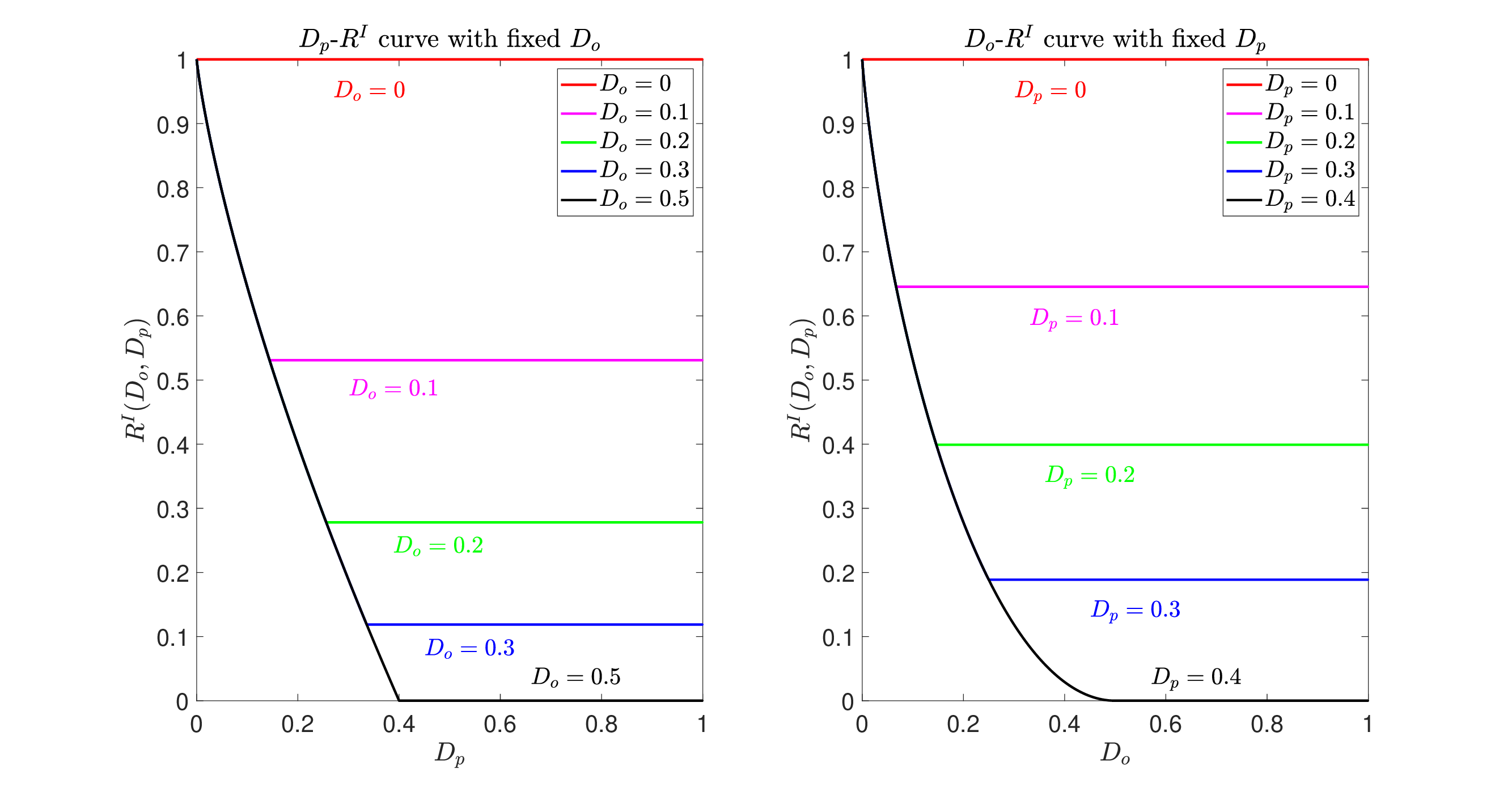}}
\caption{$R(D_p,D_o)$ for $\rho=0.5$ and $q = 0.9$ with One Variable Fixed in the Binary Case.}
\label{fixed-one}
\end{figure}

$R(D_p,D_o)$ curves $(\rho = 0.5, q=0.9)$ with one variable fixed is illustrated in Figure \ref{fixed-one}, where a threshold effect can be observed. For each fixed $D_o$, $R(D_p,D_o)$ decreases monotonically as $D_p$ increases. The function becomes constant when $D_p$ exceeds a threshold (determined by $a(D_o)$) while the curves coincide for different $D_o$ values below the threshold, attributing to that $R(D_p,D_o)$ depends only on $D_p$ now. A similar situation occurs for fixed $D_p$ when examining the $D_o$-$R$ relationship. The trade-off between the semantic probability fidelity based on the observation and the observation fidelity (at least for the binary case) is perspicuous: the stronger one dominates the rate determination.

\section{Experiment}\label{experiment}

\hspace{1em} In previous sections, we incorporated the constraint limiting the divergence between conditional probabilities $p_{S|X}$ and $p_{S|Y}$ into the traditional rate-distortion theory framework. This was done to characterize the semantic rate-distortion theory, and we analytically derived the compression limit. In this section, we conduct simulation experiments to substantiate the rationale behind constraining the divergence between $p_{S|X}$ and $p_{S|Y}$ in semantic communication, which means that incorporating this constraint significantly aids the accurate transmission of information at the semantic level and bit rates savings. It is important to note that we are not aiming to propose a new specific source coding scheme. The exclusive objective of performing and presenting these experiments is to demonstrate the critical role that constraining the divergence between $p_{S|X}$ and $p_{S|Y}$ plays in semantic communication.

\subsection{Experiment Procedure}
\hspace{1em} The MNIST dataset serves as the experimental platform for this exploration. We adopt an Autoencoder (AE) framework, jointly training the encoder and decoder in a manner similar to the setup in \cite{14}. Both encoder $f_e$ and decoder $g_e$ are implemented as deep neural networks (DNNs). Encoder $f_e$ processes input $x$ through a multilayer perceptron (MLP), mapping it to a $d$-dimensional latent vector $h$. This latent vector is then quantized component-wise into $L$ levels, yielding codeword $w_s$. Upon receiving $w_s$, decoder $g_e$---also structured as a MLP---reconstructs output $y = g_e(w_s)$.

To ensure differentiability in quantization, we employ relaxation techniques following \cite{MATTG2018,14}. For simplicity, we follow the convention from \cite{14}, treating rate $R_e$ as $ d \log L$ in subsequent analysis. While $d \log L$ only provides an upper bound on the coding rate (since entropy $H(w_s) \leq d \log L$), prior work \cite{ATMTV2019} demonstrates that the actual rate closely approaches this bound.

The training and evaluating stages are as follows.

\begin{itemize}
    \item Train the encoder and decoder jointly with the training set.

Each handwritten digit image $x_{tr}$ from the training set is sequentially processed through semantic encoder $f_e$ and decoder $g_e$, undergoing lossy compression and reconstruction at fixed rate $R_e$ to yield output $y_{tr}$, which is then fed into a pre-trained classifier to identify the handwritten digit, producing a probability distribution as the preliminary recognition result (i.e. the semantic distribution given recovered image $y_{tr}$, namely $p_{S|y_{tr}}$). 

The Kullback-Leibler (KL) Divergence is computed between the image's original digit label, which is represented as a single-point distribution, i.e. the true semantic distribution given original image $x_{tr}$, namely $p_{S|x_{tr}}$, and distribution $p_{S|y_{tr}}$ to control the semantic distortion. The Mean-Squared Error (MSE) is computed between the original and the recovered images, $x_{tr}$ and $y_{tr}$, to control the symbolic distortion. Semantic encoder $f_e$ and decoder $g_e$ are jointly trained minimizing the composite objective,
\begin{equation}
    \mathcal{L}=\underbrace{d_{MSE}(x_{tr},y_{tr})}_{Symbolic\ distortion}+\gamma\cdot \underbrace{D_{KL}(p_{S|x_{tr}}||\ p_{S|y_{tr}})}_{Semantic\ distortion}.
\end{equation}

\begin{figure}[t]
\centerline{\includegraphics[width=1\textwidth,trim = 0 0 0 0, clip]{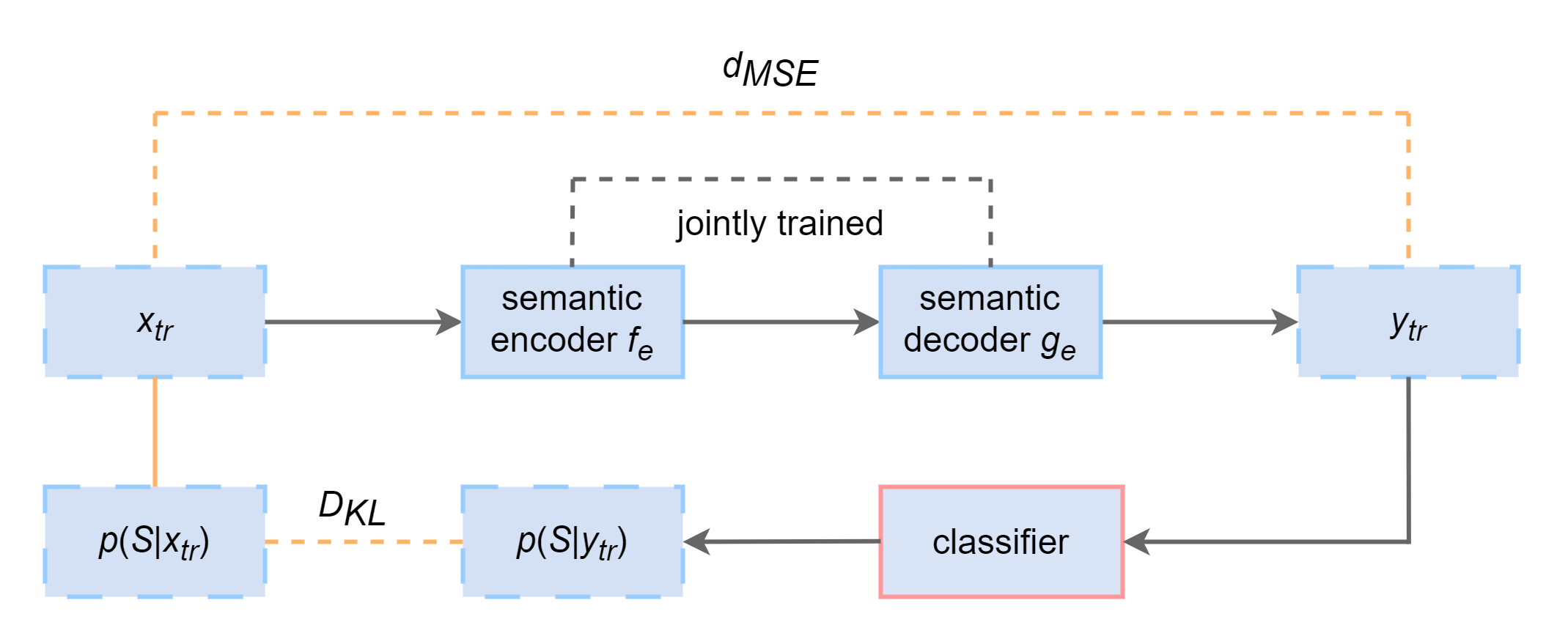}}
\caption{Training Workflow.}
\label{TR}
\end{figure}

Figure 4 illustrates this training workflow, where $\gamma$ is tunable. $\gamma=0$ corresponds to the traditional communication that ignores semantics while $\gamma > 0$ implements the semantic communication with varying degrees of semantics preservation.

    \item Evaluate the encoder and decoder on the test set.
    
For evaluating, each handwritten digit image $x_{te}$ from the test set is processed through pre-trained semantic encoder $f_e$ and decoder $g_e$ (optimized in Training Phase at fixed rate $R_e$). Output $y _{te}$ is subsequently fed into the same classifier employed during training for digit recognition, with a probability distribution generated. The final recognition result is determined by selecting the digit with the maximum probability.

We compare all recognition results against ground-truth digit labels to calculate the recognition accuracy. Crucially, the accuracy reflects traditional communication performance for $\gamma=0$ while the accuracy quantifies semantic communication efficacy at distinct semantic-awareness levels for $\gamma > 0$.

The evaluating workflow is illustrated in Figure 5.

\begin{figure}[t]
\centerline{\includegraphics[width=1\textwidth,trim = 0 0 0 0, clip]{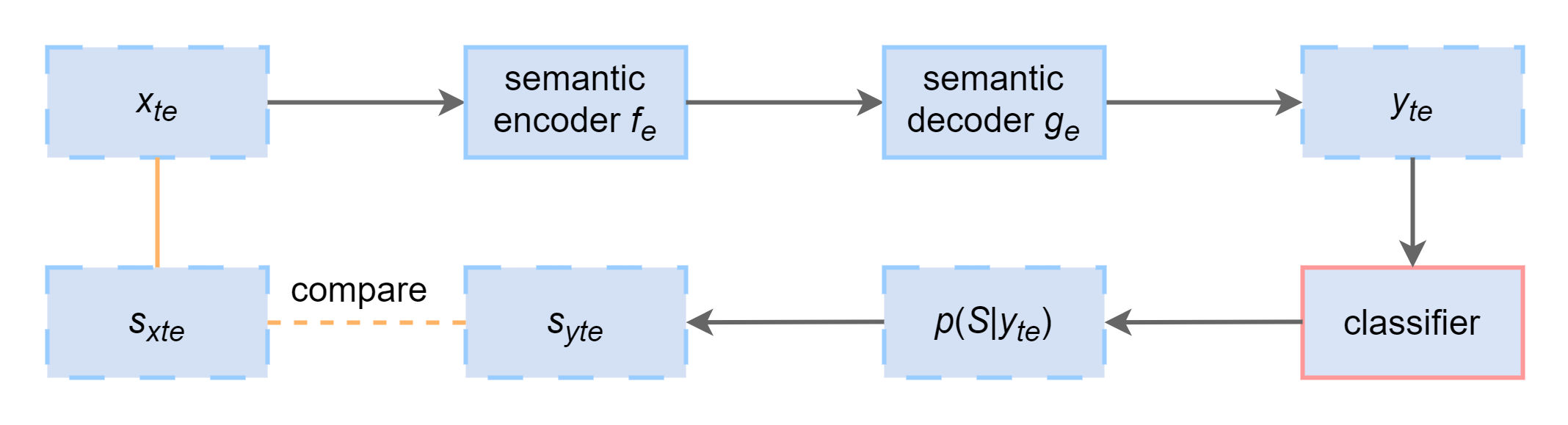}}
\caption{Evaluating Workflow.}
\label{TE}
\end{figure}

\end{itemize}

\subsection{Experiment Results}
\hspace{1em} We train and evaluate encoders and decoders across varying bit rates $R_e$ and $\gamma$ values, recording comprehensive results in Table \ref{table1}. These data strongly validate the rationality and necessity of constraining distortion between conditional probabilities $p_{S|y}$ and $p_{S|x}$ in semantic communication systems.

\begin{itemize}
    \item \textbf{Enhancing the semantic accuracy.}
    The constraint \textbf{improves semantic-level information preservation} to a great extent. At $R_e=12$ bits, traditional communication ($\gamma=0$) achieves only 62.41\% recognition accuracy after lossy compression and reconstruction. By contrast, introducing a minimal-weight constraint ($\gamma=0.01$) on the conditional probability distortion elevates accuracy beyond 90\% without additional bit allocation. Table 1 and Figure 7 reveal that tightening the conditional probability distortion requirement (increasing $\gamma$) in a way consistently boosts handwritten digit recognition accuracy at equivalent bit rates, confirming its critical role in semantic information protection. In addition, as shown in Figure \ref{10}, in traditional communication at $R_e=10$ bits ($\gamma$=0), the limited number of bits representing the compressed image results in insufficient clarity of the recovered one. This leads to frequent misclassification cases (e.g., the two ``4" digits in the first row are easily mistaken for ``9", and the ``8" in the third row is often misidentified as ``3"), consequently causing low recognition accuracy. However, after introducing the constraint on conditional semantic probability distortion (e.g., $\gamma$=0.01) at the same 10 bits rate, misclassification instances significantly decrease. This demonstrates that conditional semantic probability distribution distortion constraint \textbf{helps ensure semantic stability}.

    \item \textbf{Optimizing the bit-rate efficiency.}
    The constraint reduces bandwidth demand for target accuracy thresholds. For 40\% semantic accuracy, the traditional way ($\gamma=0$) requires more than 6 bits while semantic communication ($\gamma=0.1$) achieves 40\% accuracy at just 2 bits. If the goal is to attain accuracy above 85\%, the traditional methods ($\gamma=0$) need at least 36 bits, whereas semantic communication ($\gamma=0.1$) only asks 4 bits. These empirical evidence establishes that limiting conditional probability distortion compellingly saves the necessary bit rates in bandwidth-efficient transmission.
\end{itemize}

\begin{table}[ht]
\setlength{\abovecaptionskip}{0.5cm}
\setlength{\belowcaptionskip}{0.5cm} 
\begin{center}
\caption{Performance Data for Traditional and Semantic Communication.}
\label{table1}
\begin{tabular}{c|c|c|c|c|c|c} 
 L & dim & \textbf{R} & 
\textbf{$\gamma$} & $d_{MSE}(x_{te},y_{te})$ &$D_{KL}(p_{S|x_{te}}||\ p_{S|y_{te}})$ & \textbf{Accuracy} \\
  \hline
2	&2  &2	&0	&0.0577	&2.2810	&0.0892\\
2	&2	&2	&0.01	&0.0583	&2.1724	&0.2056\\
2	&2	&2	&0.1	&0.0767	&1.3595	&0.4000\\
2	&4	&4	&0	&0.0476	&2.2334	&0.1461\\
2	&4	&4	&0.01	&0.0511	&1.3455	&0.7796\\
2	&4	&4	&0.1	&0.0707	&0.3552	&0.9502\\	
4	&3	&6	&0	&0.0436	&2.1969	&0.3523\\
4	&3	&6	&0.01	&0.0479	&1.1260	&0.8339\\
4	&3	&6	&0.1	&0.0726	&0.2411	&0.9703\\
4	&4	&8	&0	&0.0379	&2.1314 &0.4947\\
4	&4	&8	&0.01	&0.0430	&0.9271 &0.9016\\
4	&4	&8	&0.1	&0.0651	&0.2044	&0.9763\\
4	&5	&10	&0	&0.0337	&2.0647	&0.5994\\
4	&5	&10	&0.01	&0.0387	&0.8314	&0.9357\\
4	&5	&10	&0.1	&0.0596	&0.1852	&0.9787\\
4	&6	&12	&0	&0.0305	&2.0389	&0.6241\\
4	&6	&12	&0.01	&0.0353	&0.8343	&0.9419\\
4	&6	&12	&0.1	&0.0571	&0.1743	&0.9816\\
8	&6	&18	&0	&0.0297	&1.8821	&0.7171\\
8	&8	&24	&0	&0.0247	&1.7237	&0.7880\\
8	&10	&30	&0	&0.0211	&1.6614	&0.8313\\
8	&12	&36	&0	&0.0189	&1.5860	&0.8496\\

\end{tabular}
\end{center}
\end{table}

\begin{figure}[H]
\centerline{\includegraphics[width=1\textwidth,trim = 0 0 0 0, clip]{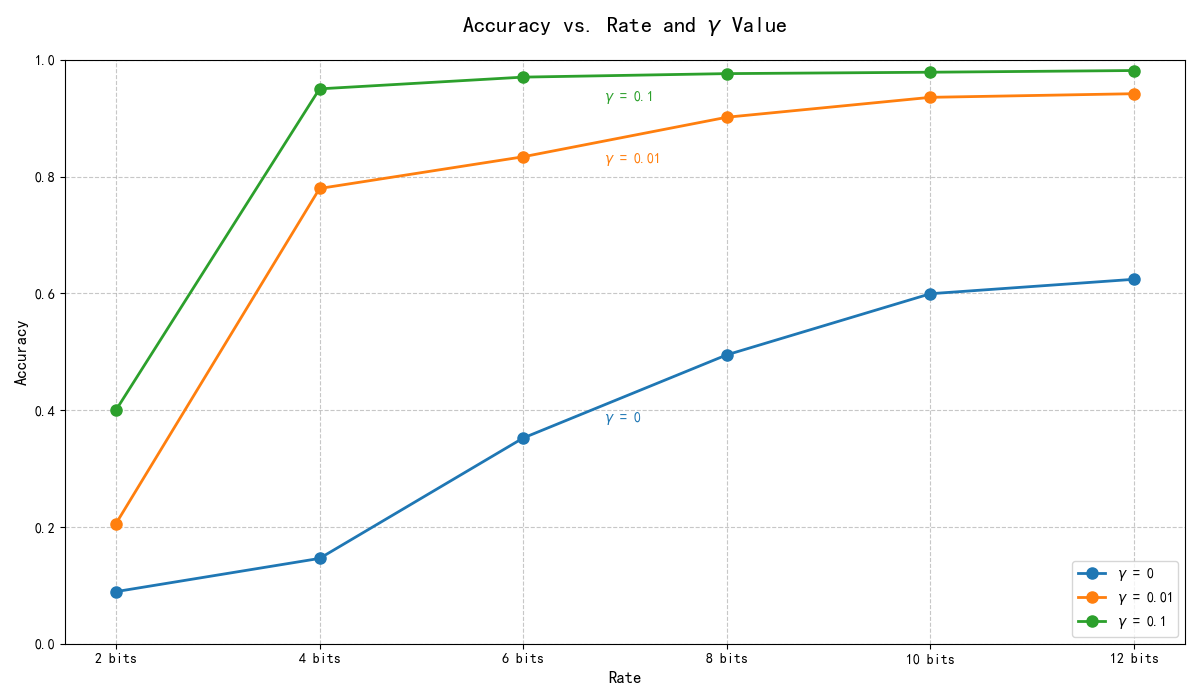}}
\caption{Accuracy under Different Rates and $\gamma$ Values.}
\label{ZX}
\end{figure}

\begin{figure}[H]
\centerline{\includegraphics[width=0.9\textwidth,trim = 0 0 0 0, clip]{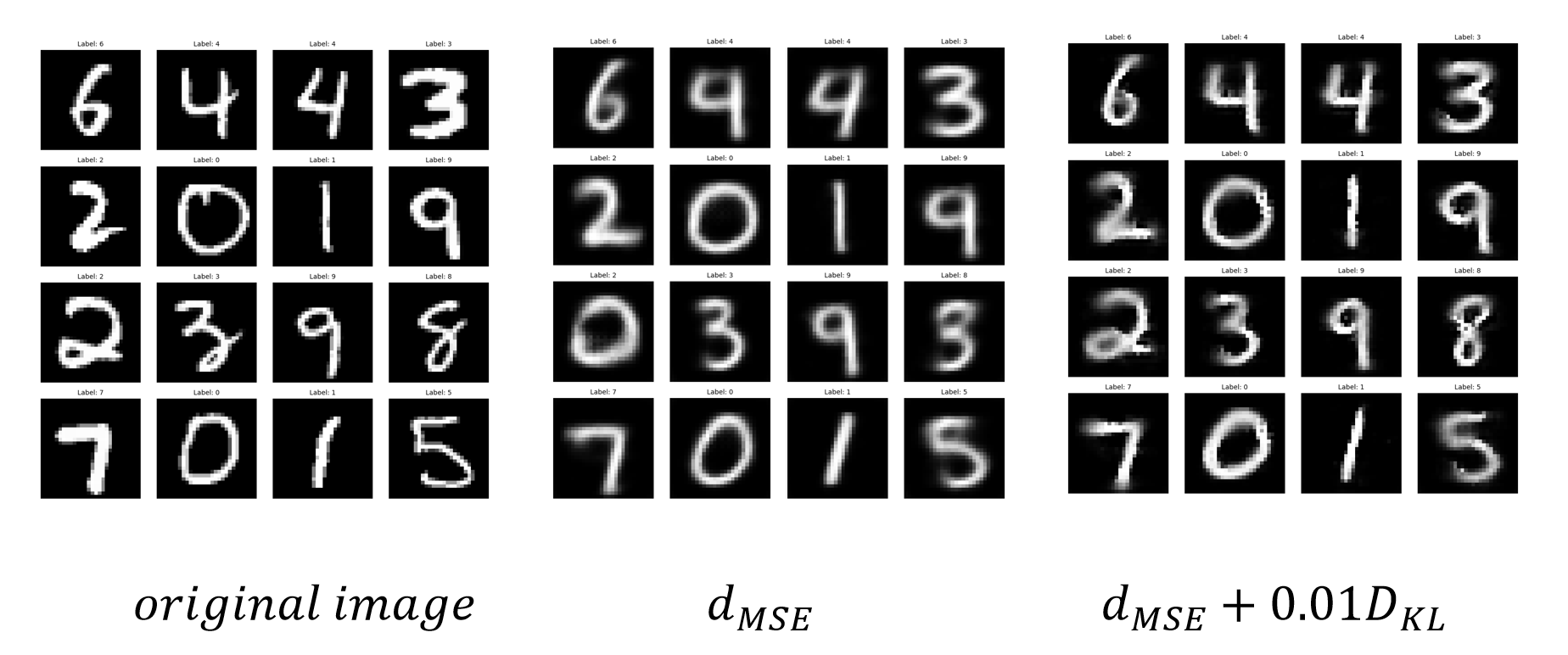}}
\caption{The Original and Recovered Images for Traditional and Semantic Communication at 10 Bits.}
\label{10}
\end{figure}

\begin{table}[ht]
\setlength{\abovecaptionskip}{0.5cm}
\setlength{\belowcaptionskip}{0.5cm} 
\begin{center}
\caption{Performance Data for Traditional and Semantic Communication at 8 bits.}
\
\label{table2}
\begin{tabular}{c|c|c|c} 
  \textbf{$\gamma$} & $d_{MSE}(x_{te},y_{te})$ & $D_{KL}(p_{S|x_{te}}||\ p_{S|y_{te}})$&\textbf{Accuracy} \\
  \hline
0&	0.0379&	2.1314&	0.4947\\
0.01&	0.0430&	0.9271&	0.9016\\							
0.5&	0.0908&	0.1206&		0.9788\\
100&	0.2769&	0.1010&		0.9791\\

\end{tabular}
\end{center}
\end{table}

\begin{figure}[H]
\centerline{\includegraphics[width=1\textwidth,trim = 0 0 0 0, clip]{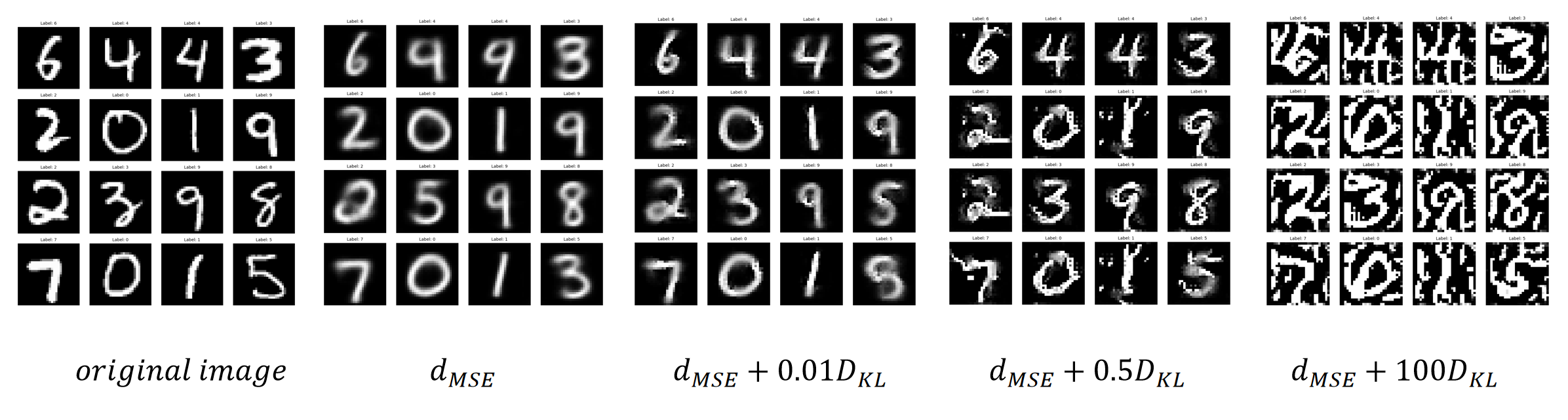}}
\caption{The Original and Recovered Images for Traditional and Semantic Communication at 8 Bits.}
\label{8R}
\end{figure}

Figure \ref{8R} contrasts the original against recovered images at $R_e=8$ bits under traditional communication ($\gamma=0$) and semantic communication ($\gamma>0$). Table \ref{table2} provides corresponding experimental metrics. When maintaining this fixed bit rate, semantic communication yields recovered images exhibiting reduced pixel similarity to the original yet enhanced semantics preservation. Notably, stylistic variations of the same digit(e.g., ``4'', ``3'', ``9'') in the original image converge to nearly identical forms in the recovered one ($\gamma=0.5$). This phenomenon attests to that our conditional probability constraint directs the semantic encoder and decoder to retain only task-related semantic features (digit semantics) while discarding symbol-level attributes irrelevant to handwritten digit recognition --- which aligns perfectly with the essential of semantic communication.

Further intensifying the constraint to $\gamma=100$ diminishes the MSE distortion term’s influence, effectively optimizing exclusively for semantic transmission. The resulting recovered image becomes barely recognizable to human observers but still sustains beyond 97\% recognition accuracy, exemplifying that the system preserves only machine-interpretable semantic features tailored to the recognition task. This extreme case underscores the core role of constraining conditional probability distortion in task-driven semantic communication.

For real-world scenarios requiring dual preservation of semantic and symbolic information (e.g., forensic videos archiving), reducing $\gamma$ achieves balanced fidelity. As evidenced in Figure \ref{8R} and Table \ref{table2}, the $\gamma=0.01$ recovered image retains stylistic nuances of digits (e.g., distinct ``4'', ``3'',``2'' forms) while boosting recognition accuracy far beyond the one in traditional communication. This demonstrates our ability to proportionally regulate observation distortion (symbolic fidelity) and semantic probability distortion (semantic fidelity) --- precisely the key technology for AI-augmented communication systems.

\section{Conclusion}\label{conclusion}

\hspace{1em} Setting out to the fundamental distinction between semantic communication and traditional communication paradigms --- the prioritization of accurate transmission of semantic probabilities based on the symbolic strings, we proposed a realistic semantic communication model and semantic compression framework, and developed the semantic rate-distortion theory. We defined the information and operation semantic rate distortion functions separately, and have proven the equivalence of the two under the condition of lower semicontinuity, which elucidates that the information semantic rate distortion function is the minimum rate required to transmit semantic probability distributions with the same finite distortion. To get some insight, we explicitly computed the closed-form expression of the semantic rate distortion function for a binary case, with emphasis on the implication that may stimulate deeper thinking. Our experiments conclusively demonstrated that constraining the divergence between conditional semantic probabilities significantly enhances semantic transmission accuracy and optimizes bit-rate efficiency in semantic communication.

This research aims to bridge information theory with artificial intelligence, paving the way for semantic-aware communication systems. Future works will be devoted to in-depth exploration to AI-driven semantic architectures and theories on next-generation intelligent communication.

\begin{appendices}
\section{Proofs of Theorem \ref{THA}}\label{appendix A}
 \hspace{1em} It is enough to show $R^I({D}_{p},{{D}_{o}})$ itself is achievable. For each $n$, we will use the Poisson representation lemma \cite{proof} to construct the desired code. For any $\epsilon>0$, we can find a $p_{Y|X}$ such that
\begin{align}
&I(X;Y) \leq R^I({D}_{p},{{D}_{o}}) + \epsilon,\\
&Ed_o(X,Y) \leq D_o,\\
&Ed_p(p_{S|X},p_{S|Y})\leq D_p.
\end{align}
Let $p_{X^n,Y^n}$ be the $n$-times product of $p_{Y|X}p_X$  (that is, if $\{(X_i,Y_i)\}_{i=1}^n {i.i.d.} \ {\sim}\ \\p_{Y|X}p_{X}$, then $(X^n,Y^n)\sim p_{X^n,Y^n}$).
Let $\{\widetilde{Y}_i\}_{i=1}^\infty {i.i.d.} \ {\sim} \ p_{Y^n}$, $\{T_i\}_{i=1}^\infty$ be a Poisson point process, and $K = k(X^n,\{T_i\}_{i=1}^\infty,\{\widetilde{Y}_i\}_{i=1}^\infty)$ be defined by
\begin{equation}
k(x^n,\{t_i\}_{i=1}^\infty,\{\tilde{y}_i\}_{i=1}^\infty) = \arg\min_{i} t_i \frac{dp_{Y^n}}{dp_{Y^n|X^n}(\cdot|x^n)}(\tilde{y}_i).
\end{equation} 
Then by the Poisson representation lemma \cite{proof} we know that
\begin{align}
&(X^n,\widetilde{Y}_{K}) \sim p_{X^n,Y^n},
\\
&\begin{aligned}H(K) &\leq I(X^n;Y^n) + \log(I(X^n;Y^n) + 1) + 4 \\
&= nI(X;Y) + \log(nI(X;Y) + 1) + 4.\end{aligned}
\end{align}
Now define
\begin{align}
& U_{1,n} = (\{T_i\}_{i=1}^\infty,\{\widetilde{Y}_i\}_{i=1}^\infty),\label{good code 1}\\
&U_{2,n} =  \{\widetilde{Y}_i\}_{i=1}^\infty,\label{good code 2}\\
& f_n(x^n,u_{1,n}) = k(x^n , \{t_i\}_{i=1}^\infty,\{\tilde{y}_i\}_{i=1}^\infty),\label{good code 3}\\
& g_n(k,u_{2,n}) = \tilde{y}_{k}\label{good code 4},
\end{align}
then $f_n(X^n,U_{1,n}) = K$ and $g_n(K,U_{2,n}) = \widetilde{Y}_K$. 
We have
\begin{align}
& E{d}_{o}(X^n,Y^n)=\max_{i\in\{1,\dots ,n\}} E{d_o(X_i,Y_i)} = E{d_o(X,Y)} \leq D_o, \forall n,\\
& E{d}_{p}(p_{S^n|x^n},p_{S^n|y^n})=\max_{i\in\{1,\dots ,n\}} E{d_p(p_{S_i|X_i},p_{S_i|Y_i})} =E{d_p(p_{S|X},p_{S|Y})}\leq D_p, \forall n,
\end{align}
and
\begin{equation}
\begin{aligned}
\frac{H(f_n(X^n,U_{1,n}))}{n} = \frac{H(K)}{n} &= I(X;Y) + \frac{\log(nI(X;Y) + 1) + 4}{n}\\
&\leq R^I({D}_{p},{{D}_{o}}) + \frac{\log(nR^I({D}_{p},{{D}_{o}}) + n\epsilon + 1) + 4}{n} \\
&\xrightarrow{n\rightarrow\infty} R^I({D}_{p},{{D}_{o}}).
\end{aligned}
\end{equation}
Hence the semantic code and randomness $f_n$, $g_n$, $U_{1,n}$, $U_{2,n}$ satisfy the constraints, which implies the rate $R^I({D}_{p},{{D}_{o}})$ is achievable. Note that although $U_{1,n}$ and $U_{2,n}$ are not real-valued, they can be encoded by a single real number since $\mathbb{R}$ has the same cardinality as $\mathbb{R}^\infty$.

\section{Proofs of Theorem \ref{converse part}}\label{appendix B}
\hspace{1em} Now we show $R^O(D_p,D_o) \geq  R^I(D_p,D_o)$ if $ R^I(D_p,D_o)$ is lower semicontinuous. 
Suppose $(R,D_p,D_o)$ is achievable, then there exists a sequence of semantic rate distortion codes $\{f_n,g_n\}$ and a sequence of random variables $\{U_{1,n},U_{2,n}\}$ such that \eqref{achievable 1}\eqref{achievable 2}\eqref{achievable 3} hold.
Let $W_n = f_n(X^n,U_{1,n})$ and $Y^n = g_n(W_n,U_{2,n})$. Denote
\begin{align}
&P_{i,n} =  E{d_p(p_{S_i|X_i},p_{S_i|Y_i})},\ P_n = \max\limits_{i\in \{1,\dots ,n\}} P_{i,n},\\
&D_{i,n} =  E{d_o(X_i,Y_i)},\ D_n = \max\limits_{i\in\{1,\dots ,n \} }D_{i,n}.
\end{align}
Then
\begin{equation}
\begin{aligned}
R &\geq \limsup\limits_{n\rightarrow\infty} \frac{H(W_n)}{n} \\
& \geq \limsup\limits_{n\rightarrow\infty} \frac{I(X^n;W_n)}{n}\\
&\geq \limsup\limits_{n\rightarrow\infty} \frac{I(X^n;Y^n)}{n}\\
&= \limsup\limits_{n\rightarrow\infty} \frac{1}{n}(H(X^n) - H(X^n|Y^n))\\
&= \limsup\limits_{n\rightarrow\infty} \frac{1}{n}\sum\limits_{i=1}^n (H(X_i)-H(X_i|X^{i-1},Y^n))\\
&\geq \limsup\limits_{n\rightarrow\infty} \frac{1}{n}\sum\limits_{i=1}^n (H(X_i)-H(X_i|Y_i))\\
&= \limsup\limits_{n\rightarrow\infty} \frac{1}{n}\sum\limits_{i=1}^n I(X_i;Y_i).\\
\end{aligned}
\end{equation}
And
\begin{equation}
\begin{aligned}
R
&\geq \limsup\limits_{n\rightarrow\infty} \frac{1}{n}\sum\limits_{i=1}^n R^I(P_{i,n},D_{i,n})\\
&\overset{(a)}{\geq} \limsup\limits_{n\rightarrow\infty} R^I(P_{n},D_{n}),\\
\end{aligned}
\end{equation}
where $(a)$ holds because $R^I(D_p,D_o)$ is monotonically decreasing in each variable. 
By \eqref{achievable 2} and \eqref{achievable 3}, for any $\epsilon > 0$ and all sufficiently large $n$ we have
\begin{equation}
P_n \leq D_p + \epsilon,\\
D_n \leq D_o + \epsilon.
\end{equation}
Therefore, $R \geq R^I (D_p+\epsilon,D_o+\epsilon)$. 
Since $R^I(D_p,D_o)$ is lower semicontinuous, letting $\epsilon\rightarrow0$ we obtain
\begin{equation}
R \geq \liminf\limits_{\epsilon\rightarrow0} R^I(D_p+\epsilon,D_o+\epsilon) \geq R^I(D_p,D_o).
\end{equation}
 
\section{Proofs of Proposition \ref{Lower semicontinuity}}\label{appendix C}
\hspace{1em} To show Proposition \ref{Lower semicontinuity}, we first prove the following proposition. For $x,y\in\mathbb{R}^n$ we write $x\prec y$ ($x\preceq y$) to denote that $x_k<y_k$ ($x_k\leq y_k$) for all $k\in\{1,\cdots,n\}$, and $x\succ y$ ($x\succeq y$) means $x_k>y_k$ ($x_k\geq y_k$) for all $k\in\{1,\cdots,n\}$.

\begin{proposition}\label{optimization over continuous func is still continuous}
Suppose $f:\mathbb{R}^n\rightarrow \mathbb{R}$ is a continuous function, $g:\mathbb{R}^n\rightarrow \mathbb{R}^m$ is a continuous map, and $C\subset \mathbb{R}^n$ is a compact set. For $x\in\mathbb{R}^m$, let $A_x = \{t\in C: g(t)\preceq x\}$. Define $\Omega = \{x\in\mathbb{R}^m: A_x\neq \emptyset\}$ and the function $h: \Omega\rightarrow \mathbb{R}$ as
\begin{equation}
\begin{aligned}
h(x) = \inf\{f(t):t\in C, g(t)\preceq x\}.
\end{aligned}
\end{equation}
Then $h(x)$ is a lower semi-continuous function.
\end{proposition}

\textbf{Proof: } For any $x\in\Omega$ and a sequence $x_n\in \Omega$ with $x_n\xrightarrow{n\rightarrow\infty}x$, we show that there exists a subsequence $x_{n_k}$ such that
\begin{equation}\label{eq subsequence}
\liminf\limits_{k\rightarrow\infty}h(x_{n_k}) \geq h(x).
\end{equation}
The above statement is sufficient to imply that $h(x)$ is lower semicontinuous at $x$. Suppose not, then we can find a sequence $x_n\in\Omega$ with $x\xrightarrow{n\rightarrow\infty}x$ such that
\begin{equation}
\lim\limits_{n\rightarrow \infty}h(x_n) < h(x),
\end{equation}
which is contradict to \eqref{eq subsequence}.

Now we prove \eqref{eq subsequence}. For any $\epsilon>0$ and each $n$, we can find $t_n \in A_{x_n}$ such that $f(t_n) \leq h(x_n) + \epsilon$. Since $\{t_n\}_{n=1}^\infty\subset C$ and $C$ is compact, there exists a subsequence $\{t_{n_k}\}_{k=1}^\infty$ such that $t_{n_k}\xrightarrow{k\rightarrow\infty} t_0 \in C$. In addition,  we have
\begin{equation}
g(t_{n_k}) \preceq x_{n_k}, \ \forall k,
\end{equation}
which together with the continuity of $g$ implies that $g(t_0) \preceq x$. Consequently, we have $t_0 \in A_x$ and hence $f(t_0) \geq h(x)$. Therefore, using the continuity of $f$ we obtain
\begin{equation}
\liminf\limits_{k\rightarrow\infty}h(x_{n_k}) + \epsilon \geq \liminf\limits_{k\rightarrow\infty}f(t_{n_k}) = f(t_0) \geq h(x).
\end{equation}
Finally, the desired result follows from letting $\epsilon\rightarrow0$.
\hfill $\square$

Now we continue to prove Proposition \ref{Lower semicontinuity}. Since $\mathcal{X}$ and $\mathcal{Y}$ are finite, $p_{Y|X}$ can be represented by a $|\mathcal{X}|\times|\mathcal{Y}|$ matrix $W$, where $W_{x,y} = p_{Y|X}(y|x)$. Let
\begin{equation}
\triangle = \{W\in \mathbb{R}^{|\mathcal{X}|\times|\mathcal{Y}|}: W_{x,y}\geq0, \sum\limits_{y\in\mathcal{Y}}W_{x,y} = 1,\forall x\in\mathcal{X}\}.
\end{equation}
Clearly, $\triangle$ is a compact set in $\mathbb{R}^{|\mathcal{X}||\mathcal{Y}|}$. Recall that $I(X;Y)$ is convex with respect to $p_{Y|X}$ (see \cite{Cover2006}[Theorem 2.7.4])and hence continuous on $\triangle$. Since
\begin{equation}
Ed_o(X,Y) = \sum\limits_{x\in\mathcal{X},y\in\mathcal{Y}}W_{x,y}p_X(x)d_o(x,y),
\end{equation}
which implies $Ed_o(X,Y)$ is a linear function of $W$ thus also continuous. Note that
\begin{equation}
Ed_p(p_{S|X},p_{S|Y}) =
\sum\limits_{y:p_{Y|X}\notin S_y}\sum\limits_{x\in\mathcal{X},y\in\mathcal{Y}} W_{x,y}p_X(x) d_p\left(p_{S|x},p_{S|y}\right),
\end{equation}
where
\begin{equation}
p_{S|y}(s)  = \frac{\sum\limits_{x\in\mathcal{X}}W_{x,y}p_{X,S}(x,s)}{\sum\limits_{x\in\mathcal{X}}W_{x,y}p_X(x)}.
\end{equation}
Clearly, $p_{S|Y}$ is a continuous map from $\triangle\setminus\cup_{y\in\mathcal{Y}}S_y$ to $\mathbb{R}^{|\mathcal{S}|}$. Because $d_p(\cdot,\cdot)$ is continuous with respect to the second argument, and \eqref{regular condition} holds, we conclude that $E(d_p(p_{S|X},p_{S|Y}))$ is also a continuous function on $\triangle$. Finally, by Proposition \ref{optimization over continuous func is still continuous} we know that $R^I(D,P)$ is lower semicontinuous.

\section{Proofs of Theorem \ref{binary-case-closed-form}}\label{appendix D}
\hspace{1em} Let the conditional probability $p_{Y|X}$ be parameterized as
\begin{equation}
p_{Y|X} = \begin{bmatrix} w & 1 - w\\ z & 1-z\end{bmatrix},
\end{equation}
where $w,z\in[0,1]$. Since $\rho = 0.5$ and $q_1=q_2 = q$, by some simple calculations we obtain the distributions as follows:
\begin{align}
p_X(0) &= p_X(1) = 0.5,
\end{align}
\begin{align}
    p_{S|X} &= \begin{bmatrix} q & 1-q \\ 1-q & q \end{bmatrix},
\end{align}
\begin{align}
    p_Y(0) &= \frac{z+w}{2} , p_Y(1) = \frac{2-w-z}{2},
\end{align}
\begin{align}
    p_{S|Y} &= \begin{bmatrix} \frac{qw+(1-q)z}{w+z} & \frac{(1-q)w+qz}{w+z}\\ \frac{q(1-w) + (1-q)(1-z)}{2-w-z}& \frac{(1-q)(1-w) + q(1-z)}{2-w-z}\end{bmatrix}.
\end{align}
Therefore,
\begin{equation}
\begin{aligned}
I(X;Y) &= H(Y) - H(Y|X) \\
&= h_2\left(\frac{w+z}{2}\right) - \frac{h_2(w) + h_2(z)}{2} := I(w,z),
\end{aligned}
\end{equation}
\begin{equation}
Ed_H(X,Y) = \frac{1-w}{2} + \frac{z}{2} = \frac{1+z-w}{2} := \Gamma(w,z),
\end{equation}
\begin{equation}
\begin{aligned}
&Ed_\text{TV}(p_{S|X},p_{S|Y}) \\
= &\ \sum\limits_{x,y\in\{0,1\}}p_X(x)p_{Y|X}(y|x)d_{\text{TV}}(p_{S|x},p_{S|y}) \\
= &\ \frac{w}{2}\frac{|1-2q|z}{w+z} + \frac{1-w}{2}\frac{|1-2q|(1-z)}{2-w-z} + \frac{z}{2}\frac{|1-2q|w}{w+z}+ \frac{1-z}{2}\frac{|1-2q|(1-w)}{2-w-z}\\
= &\ |1-2q|\left(\frac{wz}{w+z} + \frac{(1-w)(1-z)}{2-w-z}\right) := \Lambda(w,z).
\end{aligned}
\end{equation}
With the above expressions, the optimization problem \eqref{definition RI} can be formulated as
\begin{equation}\label{binary case parameteried RI}
R(D_p,D_o) = \min\{I(w,z): \Lambda(w,z)\leq D_p,\Gamma(w,z)\leq D_o,w,z\in[0,1]\}.
\end{equation}
To solve the closed-form of \eqref{binary case parameteried RI}, we first establish some key observations in the following propositions. Recall that $C = |1-2q|$ and $a(D_o)$ is given by \eqref{aD_o}.
\begin{proposition}\label{D_p > a(D_o) case}
Let $D_o\in[0,1]$. If $D_p > a(D_o)$ then
\begin{equation}
\Gamma(w,z) \leq D_o \Rightarrow \Lambda(w,z) \leq D_p.
\end{equation}
\end{proposition}
\textbf{proof: } If $D_o \geq 1/2$, we have $D_p > a(D_o) = C/2$. Using the harmonic mean (HM) - arithmetic mean (AM) inequality we obtain
\begin{equation}
\Lambda(w,z) = \frac{C}{2}\left(\frac{2wz}{w+z} + \frac{2(1-w)(1-z)}{2-w-z}\right)\leq \frac{C}{2}\left(\frac{w+z}{2} + \frac{2-w-z}{2}\right) = \frac{C}{2},
\end{equation}
which implies $\Lambda(w,z) \leq D_p$ immediately.

Now consider the case $D_o < 1/2$. Suppose $\Gamma(w,z) \leq D_o$ holds, this together with $w,z\in[0,1]$ implies $0\leq z\leq 2D_o$ and $z+1-2D_o \leq w \leq 1$. For each fixed $z$, consider the function $\varphi(w) = \Lambda(w,z)$. The derivative of $\varphi(w)$ is given by
\begin{equation}
\varphi'(w) = \left(\frac{z}{w+z}+\frac{1-z}{2-w-z}\right)\frac{z-w}{(w+z)(2-w-z)}.
\end{equation}
This implies $\varphi(w)$ is decreasing on the interval $[z,1]$. Since $D_o<1/2$, we have $ z+1-2D_o > z$. Therefore, the maximum of $\varphi(w)$ on the interval $[z+1-2D_o,1]$ is taken when $w = z+1-2D_o$. Consequently,
\begin{equation}
\Lambda(w,z) \leq \Lambda(z+1-2D_o,z) := \phi(z),\ \forall z\in[0,2D_o],w\in[z+1-2D_o,1].
\end{equation}
Taking derivative of $\phi(z)$ we obtain
\begin{equation}
\phi'(z) = \frac{2(1-2D_o)^2(D_o-z)}{(2z+1-2D_o)^2(1-2z+2D_o)^2},
\end{equation}
which implies the maximum of $\phi(z)$ on the interval $[0,2D_o]$ is taken at $z= D_o$. It follows that
\begin{equation}
\Lambda(w,z) \leq \phi(z) \leq \phi(D_o) = 2CD_o(1-D_o) = a(D_o) \leq D_p.
\end{equation}
\hfill $\square$


\begin{proposition}\label{key propostion for binary case}
Consider the optimization problem
\begin{equation}\label{only Dp constraint}
\mathop{argmin}_{ w,z\in[0,1]}I(w,z),\ s.t.\ \Lambda(w,z)\leq D_p.
\end{equation}
If $D_p \leq C/2$, then one solution of \eqref{only Dp constraint} is given by
\begin{equation}
w =  \frac{1+\sqrt{1-\frac{2D_p}{C}}}{2},\ z =  \frac{1-\sqrt{1-\frac{2D_p}{C}}}{2}.
\end{equation}
\end{proposition}
\textbf{Proof: }

Note that $I(w,z) = I(z,w) = I(1-z,1-w)$ and $\Lambda(w,z) = \Lambda(z,w) = \Lambda(1-z,1-w)$. Due to these symmetric properties of $I(w,z)$ and $\Lambda(w,z)$, it is sufficient to consider the optimization problem
\begin{equation}\label{equivalent 1}
\mathop{argmin}_{w,z}I(w,z),\ s.t.\ z\in[0,\frac{1}{2}],w\in[z,1-z],\Lambda(w,z)\leq D_p.
\end{equation}
For each fixed $z\in[0,1/2]$, taking derivative with respect to $w$ we obtain
\begin{equation}
\frac{\partial I(w,z)}{\partial w} = \frac{1}{2}\log\left(1 + \frac{w-z}{(w+z)(1-w)}\right),
\end{equation}
which indicates $I(w,z)$ is monotonically increasing for $w\in[z,1-z]$. Besides, in the proof of Proposition \ref{D_p > a(D_o) case} we have shown that $\Lambda(w,z)$ is monotonically decreasing for $w\in[z,1-z]$. This implies that the minimizer of \eqref{equivalent 1} must satisfies $\Lambda(w,z) = D_p$. Consequently, \eqref{equivalent 1} is equivalent to
\begin{equation}\label{AAB}
\mathop{argmin}_{w,z}I(w,z),\ s.t.\ z\in[0,\frac{1}{2}],w\in[z,1-z],\Lambda(w,z) = D_p.
\end{equation}
Let $\theta(t) = \sqrt{t(2-t)(1-2A)}/2$, where $A = D_p/C$. It is not hard to verify that the following representations satisfy the constraint in \eqref{AAB}:
\begin{equation}
w(t) = \frac{t}{2} + \theta(t),\ z(t) = \frac{t}{2} - \theta(t),\ t\in[\frac{1-2A}{1-A},1].
\end{equation}
Therefore, it is sufficient to consider the minimizer of $\tilde{I}(t)$ defined as
\begin{equation}
\tilde{I}(t) = I(w(t),z(t)) = h_2(\frac{t}{2}) - \frac{h_2(\frac{t}{2}-\theta(t))+h_2(\frac{t}{2}+\theta(t))}{2}.
\end{equation}
Finally, one can verify that $\tilde{I}(t)$ is nonincreasing over $t\in[\frac{1-2A}{1-A},1]$. Consequently, the minimizer of $\tilde{I}(t)$ is $t=1$, which corresponds to
\begin{equation}
w = \frac{1 + \sqrt{1-2A}}{2},\ z=\frac{1 - \sqrt{1-2A}}{2}.
\end{equation}
This completes the proof of Proposition \ref{key propostion for binary case}.
\hfill $\square$

Now we continue to present the proof of Theorem \ref{binary-case-closed-form}. Suppose $D_p > a(D_o)$, then by Proposition \ref{D_p > a(D_o) case}, the optimization problem \eqref{binary case parameteried RI} is equivalent to
\begin{equation}
\min\{I(w,z):\Gamma(w,z)\leq D_o,w,z\in[0,1]\}.
\end{equation}
However, this is just the rate distortion function for symmetric binary source \cite{Cover2006}. Therefore, for the case $D_p\in(a(D_o),1]$ we have
\begin{equation}
R(D_p,D_o) = 1 - h_2\left(\min\left\{D_o,\frac{1}{2}\right\}\right).
\end{equation}

Next we consider the case $D_p \in [0,a(D_o)]$. Clearly we have $D_p \leq a(D_o) \leq C/2$, then Proposition \ref{key propostion for binary case} implies that one of the minimizer of
\begin{equation}
\min\{I(w,z):\Lambda(w,z)\leq D_p,w,z\in[0,1]\}
\end{equation}
is given by
\begin{equation}
w^* =  \frac{1+\sqrt{1-\frac{2D_p}{C}}}{2},\ z^* =  \frac{1-\sqrt{1-\frac{2D_p}{C}}}{2}.
\end{equation}
Note that $D_p \leq a(D_o) \leq 2CD_o(1-D_o)$ implies
\begin{equation}
\sqrt{1 - \frac{2D_p}{C}} \geq \sqrt{1 - 4D_o(1-D_o)} = |1-2D_o|.
\end{equation}
Therefore,
\begin{equation}
\Gamma(w^*,z^*) = \frac{1 - \sqrt{1-\frac{2D_p}{C}}}{2} \leq \frac{1 - |1-2D_o|}{2} \leq D_o.
\end{equation}
This implies $(w^*,z^*)$ is also a minimizer of \eqref{binary case parameteried RI}. Consequently, for the case $D_p \in [0,a(D_o)]$ we obtain
\begin{equation}
R(D_p,D_o) = I(w^*,z^*) = 1 - h_2\left(\frac{1 - \sqrt{1 - \frac{2D_p}{C}}}{2}\right),
\end{equation}
which completes the proof.
\end{appendices}

\bibliographystyle{IEEEtran}
\bibliography{ref}
\end{document}